\documentclass[11pt]{article}
\usepackage[font=footnotesize]{caption}
\usepackage{times}
\usepackage{geometry}
\usepackage[utf8]{inputenc}
\usepackage[T1]{fontenc}
\usepackage{enumitem,amssymb}
\usepackage{ragged2e}
\usepackage{graphicx}
\usepackage{comment}
\usepackage{multicol}
\usepackage{aas_macros}

\usepackage[usenames]{xcolor} 
\definecolor{xlinkcolor}{cmyk}{1,1,0,0}
\usepackage{url}
\usepackage[
 colorlinks=true,    
 linkcolor=xlinkcolor,     
 citecolor=xlinkcolor,     
 filecolor=xlinkcolor,  
 urlcolor=xlinkcolor,      
 final=true
]{hyperref}
\usepackage[numbers,sort&compress]{natbib}
\usepackage{enumitem}
\setenumerate{itemsep=0mm}
\usepackage[range-units=brackets, range-phrase=-]{siunitx}
\DeclareSIUnit{\parsec}{pc}
\DeclareSIUnit{\Mpc}{\mega\parsec}
\DeclareSIUnit{\h}{\mathit{h}}
\DeclareSIUnit{\hPerMpc}{\h\per\Mpc}

\begin{document}
\begin{raggedright} 

\huge
Request for Information Related to High Energy Physics and Space-Based Astrophysics \hfill \\[+1em]
\textit{Combining information from multiple cosmological surveys: inference and modeling challenges} \hfill \\[+1em]
\end{raggedright}

\normalsize

\noindent {\large \bf Focus Areas:}  

\noindent $\square$ Focus Area 1: Lunar farside \\
\noindent $\square$  Focus Area 2: International Space Station \\
\noindent $\blacksquare$  Focus Area 3: Cosmology from Optical Surveys

\noindent {\large \bf Contact Information:}\\
Erminia Calabrese (Cardiff University) [\texttt{calabresee@cardiff.ac.uk}]\\
Tim Eifler (University of Arizona) [\texttt{timeifler@arizona.edu}]\\

\noindent {\large \bf Authors} (alphabetical):  \\
David Alonso (Oxford University), Erminia Calabrese (Cardiff University), Tim Eifler (University of Arizona), Giulio Fabbian (Cardiff University/CCA), Simone Ferraro (Lawrence Berkeley National Laboratory), Eric Gawiser (Rutgers), J.~Colin Hill (Columbia University/Flatiron Institute), Elisabeth Krause (University of Arizona), Mathew Madhavacheril (Perimeter Institute), An\v{z}e Slosar (Brookhaven National Laboratory), David N. Spergel (CCA) \\

\noindent {\large \bf Endorsers (Group)}:  Roman Space Telescope Science Investigation Team ``Cosmology with the High Latitude Survey''\\

\noindent {\large \bf Endorsers (Individuals)}:  Maximilian H. Abitbol (University of Oxford), Marcelo Alvarez (Lawrence Berkeley National Lab), Jonathan Aumont (IRAP), Susanna Azzoni (University of Oxford / Kavli IPMU), Carlo Baccigalupi (SISSA), Mario Ballardini (University of Bologna), Nicola Bartolo (University of Padova), Ritoban Basu Thakur (California  Institute  of  Technology), Nicholas Battaglia (Cornell University), Rachel Bean (Cornell University), Karim Benabed (Institut d'astrophysique de Paris - Sorbonne Université/CNRS), Federico Bianchini (Stanford/SLAC), Julian Borrill (LBNL \& UC Berkeley), Thejs Brinckmann (Stony Brook University), Anthony Challinor (University of Cambridge), Susa E. Clark (Institute for Advanced Study), Asantha Cooray (UC Irvine), Gabriele Coppi (Department of Physics, University of Milano-Bicocca), Jacques Delabrouille (CNRS, laboratoire APC), Eleonora Di Valentino (Durham University), Bradley Dober (NIST / CU Boulder), Olivier Doré (JPL/Caltech), Cyrille Doux (University of Pennsylvania), Adriaan Duivenvoorden (Princeton University, Department of Physics), Jo Dunkley (Princeton University), Cora Dvorkin (Harvard University), Josquin Errard (APC / CNRS),Katherine Freese (University of Texas, Austin),Nicholas Galitzki (University of California San Diego), Kenneth Ganga (APC/CNRS/France), Ricardo Génova-Santos (Instituto de Astrofísica de Canarias), Martina Gerbino (INFN Ferrara), Vera Gluscevic (University of Southern California), Jon  E. Gudmundsson (Stockholm University), Salman Habib (Argonne National Laboratory), Ian Harrison (The University of Manchester), Katrin Heitmann (Argonne National Laboratory), Brandon Hensley (Princeton University), Christopher M. Hirata (The Ohio State University), Renée Hlozek (University of Toronto), Eric Huff (JPL/Caltech), John P. Hughes (Rutgers University), Andrew Jaffe (Imperial College London), Alejandro Jiménez Muñoz (LPSC, CNRS, Grenoble, France), Arthur Kosowsky (University of Pittsburgh, Massimiliano Lattanzi (Istituto Nazionale di Fisica Nucleare (IT)), Pablo Lemos (University of Sussex, University College London), Zack Li (Princeton University), Jia Liu (UC Berkeley), Thibaut Louis (IJCLab), Yin-Zhe Ma (University of KwaZulu-Natal), Juan Francisco Macias-Perez (LPSC),Marina Migliaccio (Tor Vergata University of Rome, Italy), Moritz Münchmeyer (University of Wisconsin - Madison), Miren Muñoz Echeverría (Laboratoire de Physique Subatomique et Cosmologie (LPSC) Grenoble), Federico Nati (Department of Physics, University of Milano-Bicocca, Italy ), Paolo Natoli (University of Ferrara, Italy), Andrina Nicola (Princeton University), Fabio Noviello (Cardiff University), Luca Pagano (University of Ferrara), Donald Petravick (University of Illinois at Urbana-Champaign), Francesco Piacentini (Sapienza University of Rome), Alice Pisani (Princeton University), Andrés A. Plazas Malagón (Princeton University), Mathieu Remazeilles (University of Manchester), Emmanuel Schaan (Lawrence Berkeley National Laboratory), Douglas Scott (University of British Columbia), Neelima Sehgal (Stony Brook University), Blake D. Sherwin (University of Cambridge), Meir Shimon (Tel Aviv University), Suzanne Staggs (Princeton University), Radek Stompor (CNRS/IN2P3, APC Laboratory), Anja von der Linden (Stony Brook University), Benjamin Wallisch (Institute for Advanced Study \& UC San Diego), Scott Watson (Syracuse University ), Martin White (UC Berkeley),Berend Winter (University College London / MSSL), Edward J. Wollack (NASA / GSFC), Mario Zannoni (University of Milano Bicocca), Andrea Zonca (University of California San Diego)\\

\noindent {\large \bf Abstract:} \\
The tightest and most robust cosmological results of the next decade will be achieved by bringing together multiple surveys of the Universe. This endeavor has to happen across multiple layers of the data processing and analysis, e.g., enhancements are expected from combining Euclid, Rubin, and Roman (as well as other surveys) not only at the level of joint processing and catalog combination, but also during the post-catalog parts of the analysis such as the cosmological inference process. While every experiment builds their own analysis and inference framework and creates their own set of simulations, cross-survey work that homogenizes these efforts, exchanges information from numerical simulations, and coordinates details in the modeling of astrophysical and observational systematics of the corresponding datasets is crucial.

\clearpage

\noindent

\section*{A decade of optical and millimeter-wave surveys of the Universe to be combined} 
Our current best understanding of the origin, composition, and evolution of the Universe is drawn from observations of the sky at different wavelengths collected by different kinds of experiments. For the last few billion years of cosmic time, as the large-scale structure (LSS) of the Universe becomes more established, information encoded in the spatial distribution, clustering, and shape distortion of galaxies is measured with optical imaging and spectroscopic redshift surveys. The very early phases are instead anchored with high-precision, high-resolution surveys of the Cosmic Microwave Background (CMB). As the CMB photons travel through the emerging and evolving LSS, multiple physical processes such as gravitational lensing, thermal and kinetic Sunyaev-Zel'dovich effect correlate the two. All these observations probe the same underlying matter and energy distribution of the Universe and combining them provides several advantages: (i) multiple, complementary routes to elucidate the cosmological model, e.g., to establish if the standard $\Lambda$CDM model is indeed correct or whether there are other subtle physical effects from additional unknown physics, (ii) increased statistical power, (iii) help in isolating physical effects that are otherwise degenerate, and (iv) minimization of the impact of survey-specific instrumental and astrophysical systematics. 

For past and current surveys this combination has been a trivial exercise, with simple addition of probes at the likelihood level when exploring models of the Universe and constraining cosmological parameters -- a good enough approximation given the level of noise and overlap in the observed sky. The increased sensitivity and the wider scientific reach of future surveys will make this work much more challenging. 
Many new surveys are on the horizon for this decade. Each of them will work to deliver their specific science program, but combined analyses will be the pillar for the tightest and most robust cosmological results, and therefore for potential new discoveries. In practice, this will require developing new infrastructure for combining multiple surveys and consistency across simulations. 

\section*{Multi-Survey Cosmological Analyses}

Future DOE experiments and NASA/ESA space missions such as the Vera C. Rubin Observatory's Legacy Survey of Space and Time (LSST)\footnote{\url{https://www.lsst.org/}}, the Euclid satellite mission\footnote{\url{https://sci.esa.int/web/euclid/}}, and the Nancy Grace Roman Space Telescope\footnote{\url{https://roman.gsfc.nasa.gov/}} aim to unveil the physical mechanism causing the accelerated expansion of the Universe (“The Cosmic Frontier”). Possible theoretical explanations require fundamentally new physics, elevating dark energy research into one of the most exciting problems in science today. Analyzing these LSS datasets and fully exploiting their synergies requires a concerted effort across many areas of science, most prominently cosmology, astrophysics, computer science, statistics, and data analysis techniques. Expected enhancement and synergies from this combination have been highlighted in several reports~\cite{2015arXiv150107897J,2017ApJS..233...21R,2018arXiv180207216D,2019arXiv190410439C,2020arXiv200404702E,2020arXiv200810663C}.

In addition to the three exciting datasets from Rubin, Roman, and Euclid, the community will have access to data from a variety of next-generation CMB experiments, in particular the ground-based Simons Observatory (SO)\footnote{\url{https://simonsobservatory.org/}} and the DOE-led CMB-S4\footnote{\url{https://cmb-s4.org/}} endeavors which will provide new maps of the microwave sky with significant overlap in survey footprint with Rubin, Roman and Euclid. These CMB datasets are highly complementary to the LSS information, both in terms of increasing the constraining power on cosmology and in terms of mitigating critical systematics of LSS analyses~\cite{2012ApJ...759...32V,2013arXiv1311.2338D,2017PhRvD..95l3512S,2020PhRvD.101f3509C,2020JCAP...12..001S,2014PhRvD..89d3516P,2017MNRAS.469.2760M,2021arXiv210301229C}. LSS will also provide tracers to delens potential future CMB missions with NASA contribution such as LiteBIRD~\cite{2018PhRvD..97d3527M}. 

While synergies between Euclid, Rubin, and Roman at the imaging and catalog levels are an important focus area that requires funding support, this document focuses on the post-catalog parts of the analysis, specifically the details of the cosmological inference process. Currently, every experiment (Euclid, Rubin, Roman, SO, CMB-S4) builds their own analysis and inference framework and creates their own set of simulations. Each experiment also has a specific working group or team exploring CMB and LSS correlations. We propose a cross-survey work effort that homogenizes these efforts, exchanges information from numerical simulations, and coordinates details in the modeling of astrophysical and observational systematics of the corresponding datasets. Ultimately, this effort will help the community build and use a joint analysis framework that can robustly combine LSS and CMB information. 

Such a framework for combined cosmological analyses should include: (i) Plans for joint simulations: building sky models that use the same underlying assumptions; (ii) Software infrastructure to bridge and/or merge likelihood codes from different experiments: creating a platform that standardizes code comparison between different analysis codes and identifying the nodes where connection between codes is needed; (iii) Software to compute consistent theoretical predictions for all observational probes: identifying key approximations, fiducial values and priors; (iv) Prescriptions for building joint terms, e.g., covariances between probes.

This RFI was specific to Rubin, Euclid, and Roman, but the range of datasets that are important to consider extends beyond these three. We specifically mention CMB experiments in this document, however data from the Dark Energy Spectroscopic Instrument, the NASA SPHEREx explorer mission, radio surveys, and other experiments will exist as well on the relevant timescale. All these missions face similar challenges as described here (analysis frameworks and numerical simulations). The advent of so many, diverse and complementary datasets in the near future poses a fantastic opportunity for the community, however in order to fully explore the joint constraining power of these datasets a coordinated effort is necessary.

\section*{Response to Specific Questions}
\subsection*{Science Enhancements}

\textit{a. What are the key dark energy science areas that will be enhanced by these activities? }

Dark Energy as a term describes our lack of understanding of the physical concepts that underlie cosmic acceleration. As such it encompasses a wide variety of fundamental physics topics including modified gravity, neutrino physics, dark matter-dark energy coupling, early dark energy, and many more. A joint analysis is required to control the systematics budget and to increase the constraining power such that the community can discriminate between the different physical concepts that explain cosmic acceleration. Below we list the main focus areas that need to be addressed in order to fully extract the cosmological information from joint survey analyses:

\begin{itemize}
    \item Observational uncertainties: For example,  photo-z errors, shear calibration, depth variations need to be parameterized consistently across the different surveys if the datasets are combined.  
    \item Astrophysical uncertainties: For example, nonlinear modeling of the density field, baryonic physics, intrinisic alignment, galaxy bias and Halo Occupation Distribution models are key astrophysical uncertainties. Better coordination on how to model these effects and how to mitigate their uncertainties is required.  
    \item Statistical uncertainties: For example, the functional form of the likelihood and, if a multivariate Gaussian is assumed, the computation of data covariances are key uncertainties in a joint CMB-LSS analysis. 
    \item Simulated analyses - quantifying the science return as a function of analysis choices. Analysis frameworks can inform simulation campaigns, survey strategy, and ultimately the analyses about the error budget as a function of the analysis choices (scales, redshifts, galaxy samples, summary statistics).   
    \item Numerical simulations: Nonlinear modeling of the density field and exploring the statistical uncertainties mentioned above requires numerical simulations. The initial conditions of these simulations should be coordinated across all survey collaborations to enable a better comparison. 
    \item Hydrodynamic simulations: Baryonic physics, intrinsic alignment, galaxy bias and Halo Occupation Distribution models require a hydrodynamic simulation campaign that is computationally extremely expensive. In order to utilize the available computing resources most effectively this simulation campaign must be informed by the composition of the error budget of a joint analysis. In other words, the requirements for a simulation campaign will be different when analyzing data from a single survey as opposed to data from multiple surveys. Simulated cosmological likelihood analyses of multi-survey data can identify the main contributors to the overall error budget and can inform a corresponding simulation campaign. A close connection between the simulated analyses and the simulation effort is required. 
    
    \noindent See also the report from the simulation task force presented in Ref.~\cite{2020arXiv200507281B}.
\end{itemize}

\noindent \textit{What level of scientific enhancement is expected by carrying them out after the datasets are public?} 

Coordinating the combination of multiple surveys before the datasets become public will guarantee that the release of the data will be accompanied with the tightest and most robust cosmological results. Starting this work after the datasets become public will not only slow down the path to this outcome, but will also encounter additional complications due to potential conflicts between individual analysis software tools and inconsistency between simulations. All information collected through this cross-survey effort should be made public immediately and code development for the analysis frameworks should be open source and in the public domain. The code and information can be used by anyone to analyze the joint catalogs.

\noindent \textit{What additional enhancements are expected if plans are put in place in the near term to enable joint data processing and analysis of public data sets?}

The post-catalog efforts described here will use the joint catalog that could be created from Rubin, Euclid, and Roman data. Additionally, thanks to its open source nature this framework can be immediately applied to precursor data from current LSS and CMB experiments. On the simulation front, this infrastructure could provide value-added products for the cosmological community such as (i) ray-tracing for lensing observables, (ii) painting of CMB observables like the thermal and kinetic Sunyaev Zeldovich effect, (iii) Halo Occupation Distribution models for the cosmic infrared background and radio galaxies at millimeter wavelengths, and optical and infrared galaxies that mimic the populations observed by various surveys, and (iv) LSS catalogs (e.g., red sequence galaxy and galaxy cluster catalogs).

\noindent \textit{b. What is the scope of work required, as well as the opportunities and costs?}

We suggest three stages:
\begin{enumerate}
    \item \textbf{Stage 1:} Forming an exploratory team from all interested surveys that includes funded FTE and postdocs. The main task of this team is to build an analysis framework interfacing analysis codes of different surveys and quantify the precision needed in suitable simulation campaigns.
    \item \textbf{Stage 2:} Informing and engaging the leading simulation experts in the effort defined in Stage 1. The goal here is to synergize existing efforts in the individual collaborations and to help design a simulation campaign that addresses the main contributors to the error budget of a joint analysis. The cost for such a simulation campaign will be defined in Stage 1.
    \item \textbf{Stage 3:} Include the information obtained through the simulation campaign in the analysis frameworks and help the community to analyze public datasets. We of course acknowledge that any joint analysis of proprietary datasets requires MOUs between the collaborations/projects. 
\end{enumerate}
    
\noindent \textit{c. What are key obstacles, impediments, or bottlenecks to advancing development of these plans?}
\begin{itemize}
    \item Person power and time commitment. 
    \item Computing time for the simulated analyses and the simulation campaigns. It is important to point out that coordinating the individual simulation efforts within each collaboration will utilize synergies, which will make the individual computing efforts more efficient. However, a joint analysis likely has more stringent requirements (size, resolution, parameter space, systematics modeling) on a simulation campaign, which will add to the computing time that has been allocated for the individual analysis efforts.
    \item Infrastructure to build and develop complex software and to collaborate across the different survey efforts. Infrastructure to host simulations or post-processed simulation results and make them accessible to the community.
\end{itemize}

\noindent \textit{d. Are there other science topics besides dark energy that drive the requirements for joint data processing or analysis?}

Combining multiple surveys will benefit exploration of all physical processes that affect the distribution of matter and energy in the Universe, such as modified theories of gravity and the presence of massive neutrinos. For this latter case, a significant detection of the neutrino mass is only possible in a combined survey analysis (see, e.g.,~\cite{S4ScienceBook,SOoverview}).

\subsection*{Collaboration and Partnerships}

\noindent \textit{k. What cooperation or partnerships between DOE and NASA could further the scientific and technology advances?}

Collaboration between the DOE experiment collaborations and the science teams of the NASA missions is critically important as well as the connection to the numerical simulation experts. We also stress the importance of international members of the science collaborations, who should be included in such partnerships wherever possible. 

\noindent \textit{l. What mix of institutions or collaboration models could best carry out the envisioned research and/or development?}

The survey collaborations need to identify and interface the relevant experts for Stage 1. Stage 2 requires High Performance Computing resources, probably at DOE leadership computing facilities and NASA Supercomputers. It also requires highly qualified personnel that work at the interface of cosmological data analysis and HPC software development. 
Stage 3 again requires close interaction of the numerical simulation experts and the analysis framework experts. At all stages a close contact to the upstream catalog generation efforts is required such that the error budget is continuously updated. 

\noindent \textit{m. What resources, capabilities and infrastructure at DOE National Laboratories or the NASA Centers (including the Jet Propulsion Laboratory (JPL)) would be beneficial for and could accelerate or facilitate research in this topic?}

Access to high-performance computing resources, e.g., at DOE facilities (NERSC, ACLF, OCLF) or NASA facilities (Pleiades), including trained personnel that works at the interface of cosmological data analysis and numerical simulations on HPC systems, is the key. Training of said personnel will require time and and funding.

\noindent \textit{n. Are there other factors, not addressed by the questions above, which should be considered in planning HEP and APD activities in this subject area?}

The RFI was specific to Rubin, Euclid, and Roman, but the range of datasets that are important to consider is much larger. We specifically mention CMB experiments in this document, however data from the Dark Energy Spectroscopic Instrument, the NASA SPHEREx explorer mission, radio surveys, and other experiments will exist as well on the relevant timescale. All these missions will face similar problems as described here (analysis frameworks and numerical simulations) and how to best coordinate and synergize the individual efforts should be discussed as part of future HEP and APD activities. 
 
\bibliographystyle{utphys}
\bibliography{references}

\providecommand{\href}[2]{#2}\begingroup\raggedright\begin{thebibliography}{10}

\bibitem{2015arXiv150107897J}
B.~{Jain}, D.~{Spergel}, R.~{Bean}, A.~{Connolly}, I.~{Dell'antonio},
  J.~{Frieman}, E.~{Gawiser}, N.~{Gehrels}, L.~{Gladney}, K.~{Heitmann},
  G.~{Helou}, C.~{Hirata}, S.~{Ho}, {\v{Z}}.~{Ivezi{\'c}}, M.~{Jarvis},
  S.~{Kahn}, J.~{Kalirai}, A.~{Kim}, R.~{Lupton}, R.~{Mandelbaum},
  P.~{Marshall}, J.~A. {Newman}, S.~{Perlmutter}, M.~{Postman}, J.~{Rhodes},
  M.~A. {Strauss}, J.~A. {Tyson}, L.~{Walkowicz}, and W.~M. {Wood-Vasey},
  ``{The Whole is Greater than the Sum of the Parts: Optimizing the Joint
  Science Return from LSST, Euclid and WFIRST},'' {\em arXiv e-prints} (Jan.,
  2015) arXiv:1501.07897, \href{http://arxiv.org/abs/1501.07897}{{\ttfamily
  arXiv:1501.07897 [astro-ph.IM]}}.

\bibitem{2017ApJS..233...21R}
J.~{Rhodes}, R.~C. {Nichol}, {\'E}.~{Aubourg}, R.~{Bean}, D.~{Boutigny}, M.~N.
  {Bremer}, P.~{Capak}, V.~{Cardone}, B.~{Carry}, C.~J. {Conselice}, A.~J.
  {Connolly}, J.-C. {Cuillandre}, N.~A. {Hatch}, G.~{Helou}, S.~{Hemmati},
  H.~{Hildebrandt}, R.~{Hlo{\v{z}}ek}, L.~{Jones}, S.~{Kahn}, A.~{Kiessling},
  T.~{Kitching}, R.~{Lupton}, R.~{Mandelbaum}, K.~{Markovic}, P.~{Marshall},
  R.~{Massey}, B.~J. {Maughan}, P.~{Melchior}, Y.~{Mellier}, J.~A. {Newman},
  B.~{Robertson}, M.~{Sauvage}, T.~{Schrabback}, G.~P. {Smith}, M.~A.
  {Strauss}, A.~{Taylor}, and A.~{Von Der Linden}, ``{Scientific Synergy
  between LSST and Euclid},''
  \href{http://dx.doi.org/10.3847/1538-4365/aa96b0}{{\em \apjs} {\bfseries 233}
  no.~2, (Dec., 2017) 21}, \href{http://arxiv.org/abs/1710.08489}{{\ttfamily
  arXiv:1710.08489 [astro-ph.IM]}}.

\bibitem{2018arXiv180207216D}
K.~{Dawson}, J.~{Frieman}, K.~{Heitmann}, B.~{Jain}, S.~{Kahn},
  R.~{Mandelbaum}, S.~{Perlmutter}, and A.~{Slosar}, ``{Cosmic Visions Dark
  Energy: Small Projects Portfolio},'' {\em arXiv e-prints} (Feb., 2018)
  arXiv:1802.07216, \href{http://arxiv.org/abs/1802.07216}{{\ttfamily
  arXiv:1802.07216 [astro-ph.CO]}}.

\bibitem{2019arXiv190410439C}
P.~{Capak}, J.-C. {Cuillandre}, F.~{Bernardeau}, F.~{Castander}, R.~{Bowler},
  C.~{Chang}, C.~{Grillmair}, P.~{Gris}, T.~{Eifler}, C.~{Hirata}, I.~{Hook},
  B.~{Jain}, K.~{Kuijken}, M.~{Lochner}, P.~{Oesch}, S.~{Paltani}, J.~{Rhodes},
  B.~{Robertson}, D.~{Rubin}, R.~{Scaramella}, C.~{Scarlata}, D.~{Scolnic},
  J.~{Silverman}, S.~{Wachter}, Y.~{Wang}, and {The Tri-Agency Working Group},
  ``{Enhancing LSST Science with Euclid Synergy},'' {\em arXiv e-prints} (Apr.,
  2019) arXiv:1904.10439, \href{http://arxiv.org/abs/1904.10439}{{\ttfamily
  arXiv:1904.10439 [astro-ph.IM]}}.

\bibitem{2020arXiv200404702E}
T.~{Eifler}, M.~{Simet}, E.~{Krause}, C.~{Hirata}, H.-J. {Huang}, X.~{Fang},
  V.~{Miranda}, R.~{Mandelbaum}, C.~{Doux}, C.~{Heinrich}, E.~{Huff},
  H.~{Miyatake}, S.~{Hemmati}, J.~{Xu}, P.~{Rogozenski}, P.~{Capak}, A.~{Choi},
  O.~{Dore}, B.~{Jain}, M.~{Jarvis}, N.~{MacCrann}, D.~{Masters}, E.~{Rozo},
  D.~N. {Spergel}, M.~{Troxel}, A.~{von der Linden}, Y.~{Wang}, D.~H.
  {Weinberg}, L.~{Wenzl}, and H.-Y. {Wu}, ``{Cosmology with the Wide-Field
  Infrared Survey Telescope -- Synergies with the Rubin Observatory Legacy
  Survey of Space and Time},'' {\em arXiv e-prints} (Apr., 2020)
  arXiv:2004.04702, \href{http://arxiv.org/abs/2004.04702}{{\ttfamily
  arXiv:2004.04702 [astro-ph.CO]}}.

\bibitem{2020arXiv200810663C}
R.~{Chary}, G.~{Helou}, G.~{Brammer}, P.~{Capak}, A.~{Faisst}, D.~{Flynn},
  S.~{Groom}, H.~C. {Ferguson}, C.~{Grillmair}, S.~{Hemmati}, A.~{Koekemoer},
  B.~{Lee}, S.~{Malhotra}, H.~{Miyatake}, P.~{Melchior}, I.~{Momcheva},
  J.~{Newman}, J.~{Masiero}, R.~{Paladini}, A.~{Prakash}, B.~{Rusholme}, N.~R.
  {Stickley}, A.~{Smith}, W.~M. {Wood-Vasey}, and H.~I. {Teplitz}, ``{Joint
  Survey Processing of Euclid, Rubin and Roman: Final Report},'' {\em arXiv
  e-prints} (Aug., 2020) arXiv:2008.10663,
  \href{http://arxiv.org/abs/2008.10663}{{\ttfamily arXiv:2008.10663
  [astro-ph.IM]}}.

\bibitem{2012ApJ...759...32V}
A.~{Vallinotto}, ``{Using Cosmic Microwave Background Lensing to Constrain the
  Multiplicative Bias of Cosmic Shear},''
  \href{http://dx.doi.org/10.1088/0004-637X/759/1/32}{{\em \apj} {\bfseries
  759} no.~1, (Nov., 2012) 32},
  \href{http://arxiv.org/abs/1110.5339}{{\ttfamily arXiv:1110.5339
  [astro-ph.CO]}}.

\bibitem{2013arXiv1311.2338D}
S.~{Das}, J.~{Errard}, and D.~{Spergel}, ``{Can CMB Lensing Help Cosmic Shear
  Surveys?},'' {\em arXiv e-prints} (Nov., 2013) arXiv:1311.2338,
  \href{http://arxiv.org/abs/1311.2338}{{\ttfamily arXiv:1311.2338
  [astro-ph.CO]}}.

\bibitem{2017PhRvD..95l3512S}
E.~{Schaan}, E.~{Krause}, T.~{Eifler}, O.~{Dor{\'e}}, H.~{Miyatake},
  J.~{Rhodes}, and D.~N. {Spergel}, ``{Looking through the same lens: Shear
  calibration for LSST, Euclid, and WFIRST with stage 4 CMB lensing},''
  \href{http://dx.doi.org/10.1103/PhysRevD.95.123512}{{\em \prd} {\bfseries 95}
  no.~12, (June, 2017) 123512},
  \href{http://arxiv.org/abs/1607.01761}{{\ttfamily arXiv:1607.01761
  [astro-ph.CO]}}.

\bibitem{2020PhRvD.101f3509C}
R.~{Cawthon}, ``{Effects of redshift uncertainty on cross-correlations of CMB
  lensing and galaxy surveys},''
  \href{http://dx.doi.org/10.1103/PhysRevD.101.063509}{{\em \prd} {\bfseries
  101} no.~6, (Mar., 2020) 063509},
  \href{http://arxiv.org/abs/1809.09251}{{\ttfamily arXiv:1809.09251
  [astro-ph.CO]}}.

\bibitem{2020JCAP...12..001S}
E.~{Schaan}, S.~{Ferraro}, and U.~{Seljak}, ``{Photo-z outlier self-calibration
  in weak lensing surveys},''
  \href{http://dx.doi.org/10.1088/1475-7516/2020/12/001}{{\em \jcap} {\bfseries
  2020} no.~12, (Dec., 2020) 001},
  \href{http://arxiv.org/abs/2007.12795}{{\ttfamily arXiv:2007.12795
  [astro-ph.CO]}}.

\bibitem{2014PhRvD..89d3516P}
R.~{Pearson} and O.~{Zahn}, ``{Cosmology from cross correlation of CMB lensing
  and galaxy surveys},''
  \href{http://dx.doi.org/10.1103/PhysRevD.89.043516}{{\em \prd} {\bfseries 89}
  no.~4, (Feb., 2014) 043516}, \href{http://arxiv.org/abs/1311.0905}{{\ttfamily
  arXiv:1311.0905 [astro-ph.CO]}}.

\bibitem{2017MNRAS.469.2760M}
P.~M. {Merkel} and B.~M. {Sch{\"a}fer}, ``{Parameter constraints from
  weak-lensing tomography of galaxy shapes and cosmic microwave background
  fluctuations},'' \href{http://dx.doi.org/10.1093/mnras/stx1044}{{\em \mnras}
  {\bfseries 469} no.~3, (Aug., 2017) 2760--2770},
  \href{http://arxiv.org/abs/1707.08153}{{\ttfamily arXiv:1707.08153
  [astro-ph.CO]}}.

\bibitem{2021arXiv210301229C}
S.-F. {Chen}, H.~{Lee}, and C.~{Dvorkin}, ``{Precise and Accurate Cosmology
  with CMBxLSS Power Spectra and Bispectra},'' {\em arXiv e-prints} (Mar.,
  2021) arXiv:2103.01229, \href{http://arxiv.org/abs/2103.01229}{{\ttfamily
  arXiv:2103.01229 [astro-ph.CO]}}.

\bibitem{2018PhRvD..97d3527M}
A.~{Manzotti}, ``{Future cosmic microwave background delensing with galaxy
  surveys},'' \href{http://dx.doi.org/10.1103/PhysRevD.97.043527}{{\em \prd}
  {\bfseries 97} no.~4, (Feb., 2018) 043527},
  \href{http://arxiv.org/abs/1710.11038}{{\ttfamily arXiv:1710.11038
  [astro-ph.CO]}}.

\bibitem{2020arXiv200507281B}
N.~{Battaglia}, A.~{Benson}, T.~{Eifler}, A.~{Hearin}, K.~{Heitmann}, S.~{Ho},
  A.~{Kiessling}, Z.~{Lukic}, M.~{Schneider}, E.~{Sellentin}, and J.~{Stadel},
  ``{Report from the Tri-Agency Cosmological Simulation Task Force},'' {\em
  arXiv e-prints} (May, 2020) arXiv:2005.07281,
  \href{http://arxiv.org/abs/2005.07281}{{\ttfamily arXiv:2005.07281
  [astro-ph.CO]}}.

\bibitem{S4ScienceBook}
{CMB-S4 Collaboration}, ``{CMB-S4 Science Book, First Edition},'' {\em arXiv
  e-prints} (Oct., 2016) arXiv:1610.02743,
  \href{http://arxiv.org/abs/1610.02743}{{\ttfamily arXiv:1610.02743
  [astro-ph.CO]}}.

\bibitem{SOoverview}
{Simons Observatory Collaboration}, ``{The Simons Observatory: science goals
  and forecasts},'' \href{http://dx.doi.org/10.1088/1475-7516/2019/02/056}{{\em
  \jcap} {\bfseries 2019} no.~2, (Feb., 2019) 056},
  \href{http://arxiv.org/abs/1808.07445}{{\ttfamily arXiv:1808.07445
  [astro-ph.CO]}}.

\end{thebibliography}\endgroup

\end{document}